\documentclass{iaus}
\usepackage{graphicx}

\title[H$_2$CO 6$\,$cm Masers in the Galaxy] %% give here short title %%
{A Review of H$_2$CO 6$\,$cm Masers in the Galaxy}

\author[Araya, Hofner, \& Goss]   %% give here short author list %%
{E. Araya$^{1,2}$, P. Hofner$^{1,2}$ \and W. M. Goss$^1$}

\affiliation{$^1$National Radio Astronomy Observatory, P.O.
Box 0, Socorro, NM 87801, USA\\[\affilskip]
$^2$New Mexico Institute of Mining and Technology,
Physics Department, 801 Leroy Place, Socorro, 
NM 87801, USA\break email: earaya@nrao.edu}

\pubyear{2007}
\volume{242}  %% insert here IAU Symposium No.
\pagerange{??--??}
\date{February 14, 2007, and in revised form ??}
\jname{Astrophysical Masers and their Environments}
\editors{J. Chapman \& W. Baan, eds.}
\begin{document}

\maketitle

\begin{abstract}
We present a review of the field of formaldehyde (H$_2$CO)
6$\,$cm masers in the Galaxy. Previous to our ongoing work,
H$_2$CO 6$\,$cm masers had been detected in the Galaxy only
toward three regions: NGC$\,$7538 IRS1, Sgr B2, and G29.96--0.02.
Current efforts by our group using the Very Large Array, Arecibo,
and the Green Bank Telescope have resulted in the detection
of four new H$_2$CO 6$\,$cm maser regions.
We discuss the characteristics of the known H$_2$CO masers
and the association of H$_2$CO 6$\,$cm masers with very young
regions of massive star formation. We also review the current
ideas on the pumping mechanism for H$_2$CO 6$\,$cm masers.
%% add here a maximum of 10 keywords, to be taken form the file <Keywords.txt>
\end{abstract}

\firstsection % if your document starts with a section,
              % remove some space above using this command.
\section{Introduction}

Formaldehyde (H$_2$CO) was the first organic polyatomic
molecule discovered in the interstellar medium. 
H$_2$CO is an asymmetric top 
molecule, however the asymmetry is small; the moment of 
inertia for rotation of the molecule about the `c' axis is slightly
greater than the moment of inertia about the `b' axis 
(Figure~1a). In the case of ortho-H$_2$CO 
(i.e., when the nuclear spins of the hydrogen atoms are parallel), 
the small asymmetry causes splitting of the rotational 
states into closely spaced energy levels known as 
K-doublets (Figure~1b). 
Electric dipole transitions between low energy
K-doublets ($\Delta K_c = \pm 1$, $\Delta J = 0$, 
Q-branch transitions, e.g., Townes \& Schawlow 1975) 
result in radio-frequency lines.

The first detection of H$_2$CO was reported by
Snyder et al. (1969).
They found H$_2$CO absorption in the 6$\,$cm line
(J$_{\mathrm{K_a K_c}} = 1_{11} - 1_{10}$;
$\nu_o = 4829.6596\,$MHz for the F=2-2 hyperfine
component, Tucker et al. 1971), i.e.,
the K-doublet transition from the lowest ortho-H$_2$CO
energy levels (Figure~1b). Soon after the first
detection of H$_2$CO, Palmer et al. (1969)
discovered H$_2$CO 6$\,$cm absorption 
against the 2.7$\,$K Cosmic Microwave
Background (CMB), implying an excitation temperature 
T$_{ex} < 2.7\,$K for the H$_2$CO 6$\,$cm 
K-doublet in Galactic dark clouds. The detection of H$_2$CO absorption
against the CMB (the so called {\it anomalous} absorption of
H$_2$CO) required an anti-inversion (cooling) mechanism 
that was promptly recognized to be caused by
H$_2$ collisions (Townes \& Cheung 1969, 
Garrison et al. 1975, Evans et al. 1975a,
Green et al. 1978).

Almost four decades after its discovery, the H$_2$CO
6$\,$cm line has been detected toward hundreds of 
regions in the Galaxy. H$_2$CO has been observed in {\it absorption} 
against the CMB, Galactic and extragalactic 
radio continuum sources (e.g., Rodr\'{\i}guez et al. 2006, 
Young et al. 2004, Sewi{\l}o et al. 2004b, Watson et al. 2003, 
Araya et al. 2002, Downes et al. 1980). In sharp contrast to
the ubiquitous H$_2$CO 6$\,$cm absorption line,
H$_2$CO 6$\,$cm {\it emission} is an extremely 
rare phenomenon: H$_2$CO 6$\,$cm emission
has been confirmed as megamaser emission only toward
four extragalactic objects (Araya et al. 2004a, Baan private
communication),
found as thermal emission only toward the 
Orion BN/KL region (Zuckerman et al. 1975, 
see also Araya et al. 2006b)\footnote{H$_2$CO 6$\,$cm thermal
emission was also reported toward comets Halley and
Machholz (1988j) (Snyder et al. 1989, 1990; see
however Bockel\'ee-Morvan \& Crovisier 1992).
Emission of the 2$\,$cm K-doublet is also rare 
(Mart\'{\i}n-Pintado et al. 1985, Johnston et al. 1984, 
Loren et al. 1983, Wilson et al. 1982, Evans et al. 1975b), 
and maser emission from the 2$\,$cm transition has not been observed.},
and only 7 Galactic maser regions have been reported
(Forster et al. 1980, Whiteoak \& Gardner 1983,
Pratap et al. 1994, Araya et al. 2005, 2006a,
2007 {\it in prep.}). In this article we review the 
field of H$_2$CO 6$\,$cm masers: the characteristics
of the known masers, the astrophysical environments
where the masers are found, and the current
ideas on the excitation of H$_2$CO 6$\,$cm masers.

\begin{figure}
\centerline{
\scalebox{0.55}{
\includegraphics[angle=-90]{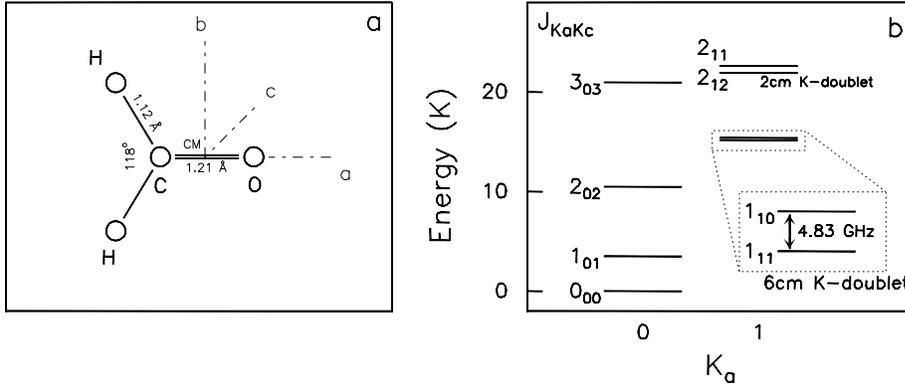}
}}
\vspace*{-0.2cm}\caption{{\it Left:} Geometry of the H$_2$CO
molecule
(Townes \& Schawlow 1975; for aesthetic reasons we show a
`left-handed' a-b-c coordinate axis). 
{\it Right:} H$_2$CO energy level diagram of 
states with E(J$_{K_aK_c}$)$<30\,$K
(Jaruschewski et al. 1986).
}\label{fig:H2CO}
\end{figure}

\vspace*{-0.3cm}
\section{H$_2$CO 6$\,$cm Maser Surveys}

The first H$_2$CO maser region detected was
NGC$\,$7538. Using the Effelsberg 100$\,$m
telescope, Downes \& Wilson (1974) detected a double
peak emission profile superimposed with an absorption
feature (see Figure~2 for a recent spectrum of the
maser). Aperture synthesis observations 
by Forster et al. (1980) and Rots et al. (1981) 
demonstrated the maser nature of the H$_2$CO
6$\,$cm emission. Approximately 10 years
after the detection of H$_2$CO emission in 
NGC$\,$7538, Whiteoak \& Gardner (1983) using the VLA
detected maser emission from five locations in 
Sgr B2. Mehringer et al. (1994) conducted 
further VLA observations ($\theta_{syn} \sim 1^{\prime\prime}$)
of the region and detected four more H$_2$CO maser
sites, resulting in a total of nine H$_2$CO 6$\,$cm
maser sites in Sgr B2. 

Since the detections by Downes \& Wilson (1974) 
and Whiteoak \& Gardner (1983), there have been 
6 surveys specifically 
focused on the search for H$_2$CO masers (see Table~1). 
Given the ubiquitous H$_2$CO 6$\,$cm absorption in 
molecular clouds (e.g., Watson et al. 2003) and the weak
intensity of the known H$_2$CO 6$\,$cm masers
(see $\S3$), surveys for H$_2$CO maser
emission have been conducted using large single
dish radio telescopes and interferometers to 
detect weak lines ($\sim 100\,$mJy) and to
avoid confusion due to H$_2$CO absorption. 

The survey by Forster et al. (1985) focused on OH maser 
sources, including not only massive 
star forming regions but also OH maser stars.
The survey yielded no new detections.
Pratap et al. (1994) and Mehringer et al.
(1995) conducted observations of active regions 
of massive star formation known to harbor
ultra-compact H{~\small II} regions as well as maser
emission from a variety of molecules. Out of 29 
sources observed with the VLA in these surveys, only
G29.96$-$0.02 was found to harbor H$_2$CO maser
emission.

Recently, Araya and collaborators conducted three surveys exploring different 
search strategies: 1. they observed regions of weak radio continuum to 
reduce confusion due to H$_2$CO 
absorption and focused on massive star forming
regions thought to be in an evolutionary stage 
prior to the ultra-compact H{~\small II} phase 
(Arecibo and GBT; Araya et al. 2004b, 2007a), 
2. they observed massive star forming regions 
independently of the radio continuum to search for 
strong and potentially variable masers (GBT;
Araya et al. 2007a), and 3. they conducted VLA
observations of sources that had been previously observed
with the GBT or Arecibo and that showed complex 
absorption line profiles consistent with H$_2$CO emission 
blended with absorption (Araya et al. 2007a, 2007 {\it in
prep.}). The three surveys resulted in detection of
four new maser regions: IRAS$\,$18566+0408 
(Araya et al. 2004b, 2005), G23.71$-$0.20 (Araya et al. 2006a, 2007a),
G23.01$-$0.41 and G25.83$-$0.18 (Araya et al. 2007 {\it in prep.}).

\begin{table}\def~{\hphantom{0}}
  \begin{center}
  \caption{H$_2$CO 6$\,$cm Galactic Masers}
  \label{tab:kd}
  \begin{tabular}{lccll}\hline
Reference                         & Sample & Telescope & Detections         & Selection Criteria      \\\hline
Downes \& Wilson (1974)           & 1      & Effelsberg& NGC$\,$7538$^*$    & NGC$\,$7538             \\
Whiteoak \& Gardner (1983)        & 1      & VLA       & Sgr B2$^*$         & Sgr B2                  \\
Forster et al. (1985)             & 19     & WSRT      & ---                & OH Maser Sources        \\
Pratap et al. (1994)              & 7      & VLA       & G29.96$-$0.02$^*$  & UCH{\small II} Regions  \\
Mehringer et al. (1995)           & 22     & VLA       & ---                & MSFR                    \\
Araya et al. (2004b)              & 15     & Arecibo   & IRAS$\,$18566+0408 & Weak Cont. MSFR         \\
Araya et al. (2007a)              & 58     & GBT/VLA   & G23.71$-$0.20$^*$  & MSFR, H$_2$CO Spectra   \\
Araya et al. (2007 {\it in prep.})& 14     & VLA       & G23.01$-$0.41 \&   & MSFR, H$_2$CO Spectra   \\
                                  &        &           & G25.83$-$0.18      &                         \\
\hline
  \end{tabular}
 \end{center}
{\scriptsize $^*$~Sources that have been observed with MERLIN and/or the
VLBA (see $\S3$).}\\
%~\vspace*{-0.5cm}
\end{table}

\vspace*{-0.3cm}
\section{Physical Properties}

H$_2$CO masers are weaker in comparison with most 
OH, H$_2$O, and CH$_3$OH masers; the flux density 
range of the known H$_2$CO masers is between 10$\,$mJy
(for the maser in IRAS$\,$18566+0408; 
see poster contribution by Araya et al. in these proceedings)
and $\sim$2$\,$Jy (for the brightest masers in Sgr B2 and NGC$\,$7538; 
Hoffman et al. 2007, Araya et al. 2007a), while most masers have flux
densities of the order of $\sim 100\,$mJy. 
Two maser regions have been observed at $\sim 50\,$mas 
angular resolution with MERLIN: NGC$\,$7538 
(Hoffman et al. 2003), and G23.71$-$0.20 (Araya et al. 
2007 {\it in prep}). MERLIN observations recover most ($\gtrsim 70\%$)
of the flux density detected at lower angular resolutions, 
and the masers are unresolved or barely resolved. 
Araya et al. (2007 {\it in prep.}) find a brightness temperature
$\gtrsim 10^5\,$K for the maser in G23.71$-$0.20.

Three sources have been observed at $\sim 10\,$mas angular
resolution with the VLBA: NGC$\,$7538, G29.96$-$0.02, and
Sgr B2. Hoffman et al. (2003) report brightness
temperatures between $10^6$ and $10^8\,$K for the 
masers in G29.96$-$0.02 and NGC$\,$7538. Two of the nine 
maser components in Sgr B2 were observed with the VLBA by
Hoffman et al. (2007). They also found brightness
temperatures in the $10^8\,$K range.
In general, only a fraction of the flux density is 
recovered with the VLBA, and as in the case of other
astrophysical masers, the lines are narrower in the 
VLBA observations compared with VLA or single dish 
observations. Hoffman et al. (2007) 
discuss these results in the context of a core-halo
model, where the maser brightness distribution
is the result of the superposition of two 
Gaussian components, one compact ($\sim 10\,$mas)
saturated component that is detected with the VLBA
(T$_b \sim 10^8\,$K), 
and one extended and unsaturated halo that is 
resolved out by the VLBA observations
(T$_b \sim 10^5\,$K).
Based on the VLBA and MERLIN results, the 
projected physical size of the masers is 
between 30 and $\sim 200\,$AU, while
the maser gain range between $-$6 and $-$12; 
the emission is unpolarized within the 
current sensitivity limits 
(Hoffman et al. 2003, 2007).

\begin{figure}
\vspace*{-3cm}\centerline{
\scalebox{0.45}{
\includegraphics[angle=-90]{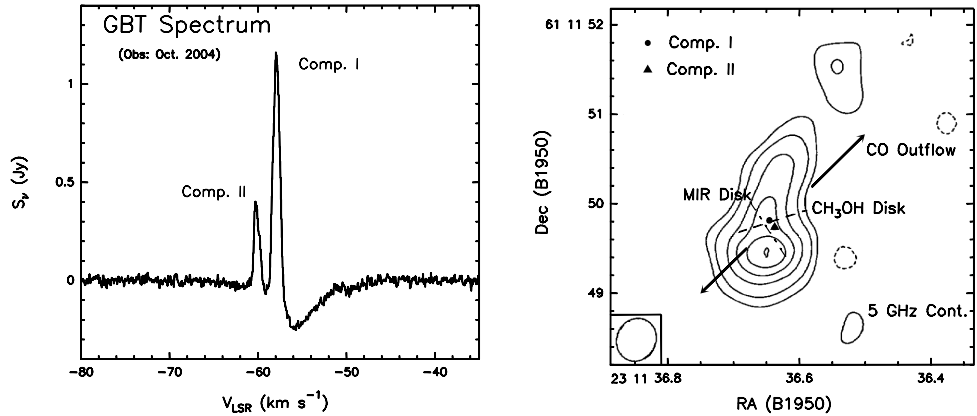}
}}
\vspace*{-0.2cm}\caption{{\it Left:} 
H$_2$CO 6$\,$cm spectrum of NGC$\,$7538 IRS1
(Araya et al. 2007a). Two maser components are 
blended with an H$_2$CO absorption line. 
{\it Right:} Location of the two H$_2$CO maser components 
(dot and triangle) superimposed on the 6$\,$cm 
radio continuum in NGC$\,$7538 IRS1 (Rots et al. 
1981). Re-reduction of the Rots et al. (1981) data
using a more accurate position of the phase calibrator
results in H$_2$CO maser positions that are consistent
with the VLBA and MERLIN 
values reported by Hoffman et al. (2003).
The direction of the CO (2--1) outflow is indicated by
arrows (Davis et al. 1998), and two possible orientations of a
circumstellar disk in the region are shown
with dot-dashed (MIR Disk, De Buizer \& Minier 2005) and
dashed (CH$_3$OH Disk, Pestalozzi et al. 2004) lines.}\label{fig:NGC7538}
\end{figure}

\vspace*{-0.3cm}
\section{Line Profiles and Velocity Gradients}

With the exception of Sgr B2 where nine H$_2$CO 
maser spots have been found (some of them showing 
multiple-peaked and broad line profiles, e.g., 
Mehringer et al. 1994), the H$_2$CO maser line 
profiles are relatively simple. 
A peculiar characteristic of the known 
H$_2$CO maser regions is that double peaked
profiles are very common; double
peaked profiles have been detected in
NGC$\,$7538 (e.g., Figure~2), G29.96$-$0.02
(Pratap et al. 1994, Hoffman et al. 2003),
IRAS$\,$18566+0408 (Araya et al. 2007c), 
G25.83$-$0.18 (Figure~3), and possibly
toward G23.01$-$0.41 (Araya et al. 2007 {\it in prep.}).
In all of these cases the velocity separation of the maser
components is less than 3$\,$km$\,$s$^{-1}$, and the 
components are (in most cases) spatially coincident
in VLA observations.
The double peaked profiles are unlikely
caused by the hyperfine structure of the
6$\,$cm H$_2$CO transition. Recent high spectral
resolution (0.1$\,$km$\,$s$^{-1}$ channel
width) VLA observations of the masers in G23.01$-$0.41 and 
G25.83$-$0.18 (Araya et al. 2007 {\it in prep.}) show
very narrow maser components (FWHM$\sim$0.3$\,$km$\,$s$^{-1}$),
possibly due to the line narrowing effect of
unsaturated masers.

In the case of the H$_2$CO masers in NGC$\,$7538, the
components are oriented in a NE--SW direction, with 
a projected separation of 79$\,$mas (240$\,$AU). 
VLBA observations of the red shifted component show
a 1900$\,$km$\,$s$^{-1}$pc$^{-1}$ velocity gradient also 
in a NE--SW orientation (see figures 4 and 5 of
Hoffman et al. 2003).

\begin{figure}
\vspace*{-2cm}\centerline{
\scalebox{0.30}{
\includegraphics[angle=0]{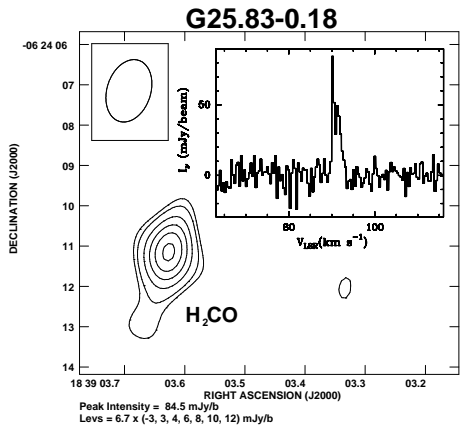}
}}
\vspace*{-0.7cm}\caption{H$_2$CO 6$\,$cm maser emission
in G25.83$-$0.18 detected with the VLA by 
Araya et al. (2007 {\it in prep.}). 
Excluding the masers in Sgr B2, five out of six sources
show double peak profiles.
}\label{fig:G25}
\end{figure}

\section{Variability}

Prior to the recent work by Araya et al. (2007c), 
variability of H$_2$CO masers had been observed
only in some of the Sgr B2 masers and in the 
NGC$\,$7538 masers; only long
time-scale variability ($> 1\,$yr) had been reported
(e.g., Mehringer et al. 1994, Hoffman et al. 2003).
In the case of the variability of the H$_2$CO 
masers in NGC$\,$7538 and based on the similar
variability-rate of the two maser 
components (Figure~4 {\it left panel}), 
Araya et al. (2007a) suggested
that the variability of the masers 
may be caused by a perturbation that took 
$\sim 14\,$yr\footnote{Curiously, the period of the 
long-term variability of H$_2$O masers in the region 
reported by Lekht et al. (2004) is $\sim 13\,$yr.}
to reach Comp. II after having reached Comp. I. 
If that were the case, an increase in the 
rate-of-change of the intensity of Comp. II would be expected
after the year $2009$. 
The variability of the two maser components could be
related to the precessing jet reported by Kraus et al. (2006).

\begin{figure}
\centerline{
\scalebox{0.45}{
\includegraphics[angle=-90]{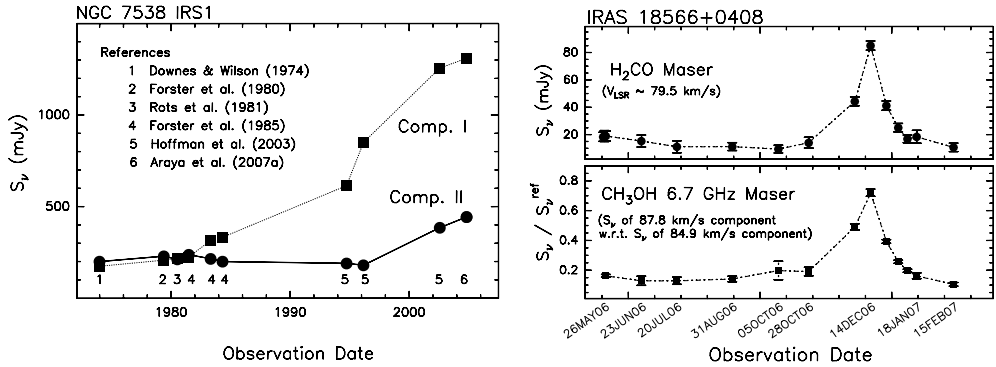}
}}
\vspace*{-4.0cm}\caption{{\it Left:} Long term 
variability of the H$_2$CO masers in 
NGC$\,$7538. The similar intensity rate-of-change of the two
maser components after the onset of the variability
lead Araya et al. (2007a) to propose that the variability
of both components may have a common origin.
{\it Right:} Arecibo light-curve of the second short-term
H$_2$CO maser flare detected (Araya et al. 2007 {\it in prep}). 
Araya and collaborators are also 
monitoring with Arecibo the CH$_3$OH 6.7$\,$GHz masers;
one of the CH$_3$OH maser components showed the same 
outburst as the H$_2$CO 6$\,$cm maser (see poster 
contribution by Araya and collaborators in 
these proceedings).}\label{fig:variability}
\end{figure}

Araya et al. (2007c) have recently found a new type of H$_2$CO 
maser variability, namely, short term flares. Using 
Arecibo, VLA, and GBT data, Araya et al. (2007c)
reported occurrence of an outburst of the 
H$_2$CO 6$\,$cm maser in IRAS$\,$18566+0408;
the maser flare lasted for less than three months and 
decayed to the pre-flare intensity within a month.
The H$_2$CO maser in IRAS$\,$18566+0408 has a double peak
profile. Both components varied by approximately the 
same factor and in the same time period; no change
in the line widths and peak velocities was detected. 
Araya et al. (2007c) discussed the implications of the
flare on the possible excitation mechanism of
the maser, and concluded that if the flare were due
to a maser gain change, then (independently
of the saturation state of the maser) the 
pumping mechanism is likely radiative; whereas
if the maser is unsaturated, then a change in the
background 6$\,$cm radio continuum might have
been amplified by the maser gas (independently
on the maser pumping mechanism).

A monitoring program of the maser with Arecibo
has recently resulted in the detection of a second
H$_2$CO maser burst in IRAS$\,$18566+0408 (Figure~4, 
{\it right panel}); showing that the flares are 
recurrent in this source. 
Araya and collaborators are 
also monitoring the CH$_3$OH 6.7$\,$GHz masers in the
region, and found that one of the CH$_3$OH maser components
showed the same outburst as the H$_2$CO 6$\,$cm 
maser (Figure~4, {\it right panel}). 
The CH$_3$OH maser peak that showed the flare does not correspond in
velocity to the H$_2$CO maser, hence the masers 
originate in different regions.
It is possible 
that both masers are unsaturated and that a change
in the background radio continuum was amplified by 
the CH$_3$OH and H$_2$CO masers 
(see poster contribution by Araya and
collaborators in these proceedings).

Variability of some of the H$_2$CO masers in Sgr B2
has also been found (Mehringer et al. 1994, Hoffman et al.
2007); however the available data are insufficient
to establish whether the masers show long term 
variability or maser flares as in IRAS$\,$18566+0408.

%\vspace*{-0.3cm}
\section{Astrophysical Environments: H$_2$CO masers pinpointing
disk candidates around young massive stars}

\begin{figure}
\centerline{
\scalebox{0.4}{
\includegraphics[angle=0]{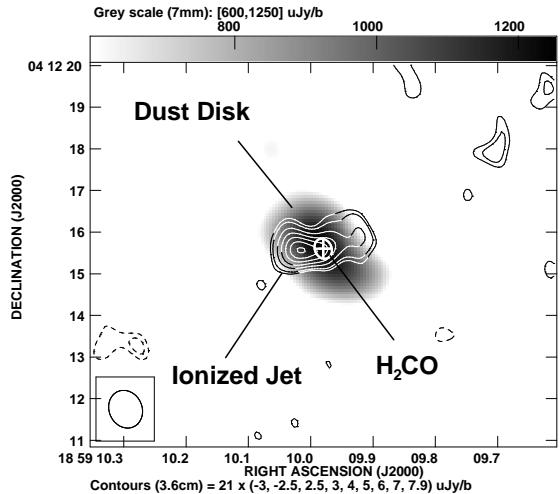}
}}
\vspace*{-0.5cm}\caption{
Location of the H$_2$CO 6$\,$cm maser in IRAS$\,$18566+0408 with respect
to 7$\,$mm (gray scale) and 3.6$\,$cm (contours) 
continuum detected with the VLA. 
Araya et al. (2007b) concluded that the 3.6$\,$cm 
emission is tracing an ionized jet whereas the 7$\,$mm 
emission is dominated by dust emission from a possible massive
circumstellar disk (torus). The H$_2$CO maser is coincident with 
the massive disk candidate. 
 }\label{fig:IR18566}
\end{figure}

Motivated by the detection of H$_2$CO maser
emission toward NGC$\,$7538 and Sgr B2 (both massive
star forming regions) the subsequent surveys for H$_2$CO maser
emission have been conducted mainly toward regions of massive
star formation (Table~1). However, H$_2$CO 6$\,$cm 
observations have also been carried out toward a 
number of non-massive star forming regions 
and no new maser has been reported 
(e.g., Araya et al. 2006b, 2003; 
Rodr\'{\i}guez et al. 2006; Young et al. 2004).
Thus, H$_2$CO 6$\,$cm masers appear to be exclusively
associated with massive star formation.  

Except for some of the H$_2$CO masers in Sgr B2, 
H$_2$CO masers are mostly found along line-of-sights 
devoid of strong compact radio continuum %(S$_\nu < 20\,$mJy; 
(though continuum regions may be found nearby, e.g., 
Pratap et al. 1994); they are located close to CLASS II CH$_3$OH and 
H$_2$O masers (in many cases coincident within
a synthesized beam), deeply embedded infrared
sources and/or other evidence of massive star formation 
such as hyper-compact H{~\small II}
regions and hot molecular cores (e.g., Araya et al. 2007b, 2006a, 2005;
Hoffman et al. 2007, 2003; Pratap et al. 1994). Thus, H$_2$CO masers
appear to trace young massive stellar objects before
the onset of a bright ultra-compact H{\small II} region. 
Moreover, in the case of three of the H$_2$CO maser 
sources that have been studied in detail, there is 
some evidence for an association between H$_2$CO
maser emission and circumstellar disks:

\newpage
\noindent {\bf -- G29.96$-$0.02 } 
is a massive star forming region
that harbors an ultra-compact H{\small II} region
and a hot molecular core (e.g., Cesaroni et al.
1994). The H$_2$CO maser is coincident with the 
hot molecular core (Pratap et al. 1994).
Hot molecular cores are believed to be an evolutionary
phase prior to the formation of an ultra-compact 
H{~\small II} region (e.g., Churchwell 2002), thus
the H$_2$CO maser pinpoints a very
young region of massive star formation.
Olmi et al. (2003) report evidence of infall
and a massive rotating disk in the hot molecular
core. Thus, the H$_2$CO 6$\,$cm emission may be 
associated with a disk around a massive young stellar
object. However, the assumption of a single
massive disk might be over-simplistic given the 
complex sub-mm morphology found by Beuther et al. 
(2007).

\noindent {\bf -- NGC$\,$7538 IRS1} is a massive star 
forming region which harbors a hyper-compact 
H{~\small II} region (e.g., Sewi{\l}o et al. 2004a),
and a CO outflow centered at the 
NGC$\,$7538 IRS1 position (Figure~2, right panel).
NGC$\,$7538 IRS1 is one of the few sources where
a circumstellar disk around a massive (proto)star
has been reported. However, the orientation
of the disk is controversial. 
Based on CH$_3$OH maser data by Minier et al. (1998, 2000, 2001), 
Pestalozzi et al. (2004) report a possible Keplerian disk
oriented $\sim$ SE--NW. However, 
based on mid-IR observations, 
De Buizer \& Minier (2005) considered that the CH$_3$OH
masers are tracing the outflow and that the disk
is oriented in a NE--SW direction (i.e., perpendicular to
the CO outflow, see Figure~2, right panel).
As mentioned in $\S$4, the two H$_2$CO maser spots 
are oriented NE--SW, and the velocity gradient of
Comp. I is also in the NE--SW direction,
i.e., perpendicular to the CO outflow and parallel
to the MIR disk. The H$_2$CO 6$\,$cm masers appear to trace material 
very close to (within 1000$\,$AU) or directly 
associated with a circumstellar disk.

\noindent{\bf -- IRAS$\,$18566+0408} was classified by Zhang (2005) 
as a massive circumstellar disk candidate. 
Based on high sensitivity and angular resolution 
6, 3.6, 1.3, and 0.7$\,$cm VLA continuum observations,  
Araya et al. (2007b) recently found supporting 
evidence for the presence of a 
massive circumstellar disk in IRAS$\,$18566+0408 (see Figure~5).
The massive disk (torus) is traced by 7$\,$mm dust emission
and has an elongation almost perpendicular to an
ionized jet traced by cm radio continuum. 
The H$_2$CO maser is coincident with the center of
the massive disk candidate.

\vspace*{-0.3cm}
\section{Pumping Mechanism of H$_2$CO 6$\,$cm Masers}

Although the number of H$_2$CO 6$\,$cm maser sources
is still small, in the past few years
significant progress in the 
characterization of H$_2$CO masers and their environments
has been made; however,
a theoretical understanding of the excitation mechanism 
of H$_2$CO masers is still lacking. Even before the detection of 
the first H$_2$CO maser, several authors mentioned and/or discussed
possible excitation mechanisms that would result in 
maser emission of the 6$\,$cm line; including 
collisional excitation with H$_2$ molecules and
electrons (Thaddeus 1972; see also Fig.14 of Evans et al. 1975a), 
and infrared pumping (Litvak 1970). However, only
Boland \& de Jong (1981) developed a specific model
to explain one of the known H$_2$CO masers (NGC$\,$7538). 
This model is based on inversion via background
radio continuum radiation; however, the model
appears to be incapable of explaining most
of the known H$_2$CO masers (Araya et al. 2007b, 
Hoffman et al. 2007, 2003, Pratap et al. 1994, Mehringer et al.
1994; see however Pratap et al. 1992). 

Besides inversion by radio continuum, other 
proposed excitation mechanisms appear to be possible:
1. Hoffman et al. (2007, 2003) and Araya et al. (2005) 
find indications that the masers could be collisionally excited in
shocked regions (see also Mart\'{\i}n-Pintado et al. 1999), 2. some H$_2$CO masers are found close to deeply embedded
infrared objects and thus infrared pumping could be possible
(Araya et al. 2006a), and 3. 
H$_2$CO masers are located close to very young massive
stellar objects where a high ionization fraction is expected,
and thus electron collision may play a role
in the pumping. 

Araya et al. (2007 {\it in prep.}) explore 
excitation of H$_2$CO masers via H$_2$ and electron collisions and
have found that collision with electrons can indeed
produce an inversion of the 6$\,$cm K-doublet (see
also Thaddeus 1972). However, preliminary results by
Araya et al. (2007 {\it in prep.}) appear to require 
long path lengths ($\sim pc$ scales) to reproduce the 
brightness temperature of the known H$_2$CO masers. Parsec-scale
path lengths of coherent velocity and homogeneous physical
conditions in massive star forming regions are unlikely.
However pumping of H$_2$CO by electron collisions appears
to be a promising mechanism to explain extragalactic megamasers
(Araya et al. 2007 {\it in prep.}, see also 
Araya et al. 2004a; Baan \& Haschick 1995).
Araya et al. (2007 {\it in prep.}) also find that including radiation trapping,
the 6$\,$cm K-doublet may be inverted
at a molecular density of $\sim 10^6\,$cm$^{-3}$
(i.e., in the transition between anomalous H$_2$CO absorption
and thermalization). However, the model
depends on accurate H$_2$(ortho/para) -- H$_2$CO 
collision rates which are not available at present (e.g., Green 1991;
see also Hoffman et al. 2003).

\vspace*{-0.3cm}
\section{Why are H$_2$CO Masers so Rare?}

In spite of a number of surveys specifically focused on the 
search for H$_2$CO masers and hundreds of sources 
for which H$_2$CO 6$\,$cm absorption studies have been conducted
(e.g., Table~1, Araya et al. 2002, Watson et al. 2003, Sewi{\l}o 
et al. 2004), H$_2$CO maser emission has been detected
only toward 7 regions (and in a total of 15 maser spots at
1$^{\prime\prime}$ resolution). 
Thus, H$_2$CO maser emission is indeed
a rare phenomenon. Reformulating the ideas
presented by Mehringer et al. (1995), Pratap et al. (1992), 
and Forster et al. (1985), H$_2$CO masers may be uncommon because:
1. they are weak in comparison with other astrophysical
masers and/or are highly beamed: the brightest known 
H$_2$CO maser is just $\sim 2\,$Jy; 2. they occur at LSR
velocities close to the systemic velocity of the 
star forming regions
and thus the maser emission is highly attenuated by 
large optical depths of the H$_2$CO absorption line: H$_2$CO
masers are typically found within $\sim 5\,$km$\,$s$^{-1}$ from the 
systemic cloud velocity as traced by H$_2$CO absorption (e.g., Figure~2,
Araya et al. 2004b) and are not found
at high velocities, indicating that H$_2$CO masers are not associated
with high velocity outflows; 
and 3. the physical conditions needed for the excitation of the masers
are very specific, and thus, short-lived in massive
star forming region environments. For example,  
the H$_2$CO masers (and H{~\small II} regions) 
in Sgr B2 are distributed in a N--S direction
suggesting that star formation in Sgr B2 was triggered by a cloud 
collision event (see Sato et al. 2000) and thus the
masers may be tracing an isochrone of the
physical conditions during the massive star formation 
process (see Gardner et al. 1986).

\vspace*{-0.3cm}
\section{Summary}

Despite the ubiquitous presence of H$_2$CO $6\,$cm absorption,
there are only 7 known H$_2$CO maser regions in the Galaxy.
Recent VLBA and MERLIN observations toward 4 sources
show brightness temperatures between $10^5 - 10^9\,$K, and physical
sizes between $30 - 200\,$AU. At least two masers show long-term variability
($>$1$\,$year), and one maser source was 
recently shown to exhibit recurrent short-term 
($<$ 3$\,$months) bursts. All known H$_2$CO $6\,$cm masers are found in
massive star forming regions, and in the case of the three sources 
that have been studied in detail (NGC$\,$7538 IRS1,
G29.96$-$0.02, and IRAS$\,$18566+0408; excluding Sgr B2) the 
H$_2$CO masers pinpoint the location of candidate disks around 
massive protostars. However, the H$_2$CO maser mechanism has to 
be clarified before the masers can be used as an astrophysical probe.

\begin{acknowledgments}

The research on H$_2$CO 6$\,$cm masers by the author (E.A.)
has greatly benefited from collaboration with
H. Linz, S. Kurtz, W. Baan, M. Sewi{\l}o, E. Churchwell,
L. Olmi, C. Watson, I. Hoffman, L.F. Rodr\'{\i}guez,
G. Garay, \& P. Palmer. 
E.A. is supported by a NRAO predoctoral fellowship. 
We thank H. Linz and I. Hoffman for comments that
improved the manuscript, and
H. Beuther for sharing information about the G29.96$-$0.02 
small scale structure based on SMA sub-mm data prior to publication.
E.A. thanks the IAU for a travel grant to attend this symposium.
\end{acknowledgments}

\end{document}